\documentclass{fundam}

\usepackage{makeidx} 
\usepackage{setspace}

\usepackage{epsfig,wrapfig}
\usepackage{amssymb,amsmath,latexsym}
\usepackage{subfigure,mathabx,graphicx}
\usepackage{url}

\input xy \xyoption{all}

\newcommand{\abb}[3]{#1 \colon #2 \rightarrow #3}

\long\def\omitthis#1{\relax}

\newcommand{\proofend}{\hfill$\blacksquare$}

\begin{document}
\setcounter{page}{1}
\issue{XXI~(2010)}

\title{Matrix Graph Grammars: Transformation of Restrictions}

\author{Pedro Pablo P\'erez Velasco\\
School of Computer Science \\
Universidad Aut\'onoma de Madrid\\    	
Ciudad Universitaria de Cantoblanco, 28049 - Madrid, Spain\\
pedro.perez{@}uam.es}

\maketitle

\runninghead{P.~P.~P\'erez}{Matrix Graph Grammars}

\begin{abstract}
  In the Matrix approach to graph transformation we represent
  \emph{simple} digraphs and rules with Boolean matrices and vectors,
  and the rewriting is expressed using Boolean operations only. In
  previous works, we developed analysis techniques enabling the study
  of the applicability of rule sequences, their independence, stated
  reachability and the minimal digraph able to fire a
  sequence. See~\cite{MGGBook} for a comprehensive
  introduction. In~\cite{MGGfundamenta}, graph constraints and
  application conditions (so-called \emph{restrictions}) have been
  studied in detail. In the present contribution we tackle the problem
  of translating post-conditions into pre-conditions and vice
  versa. Moreover, we shall see that application conditions can be
  moved along productions inside a sequence (\emph{restriction
    delocalization}). As a practical-theoretical application we show
  how application conditions allow us to perform multidigraph
  rewriting (as opposed to simple digraph rewriting) using Matrix
  Graph Grammars.
\end{abstract}


\section{Introduction}
\label{sec:intro}
 
Graph transformation~\cite{graGraBook,handbook} is becoming
increasingly popular in order to describe system behavior due to its
graphical, declarative and formal nature. For example, it has been
used to describe the operational semantics of Domain Specific Visual
Languages (DSVLs,~\cite{JVLC}), taking the advantage that it is
possible to use the concrete syntax of the DSVL in the rules which
then become more intuitive to the designer.

The main formalization of graph transformation is the so-called
algebraic approach~\cite{graGraBook}, which uses category theory in
order to express the rewriting step. Prominent examples of this approach
are the double~\cite{DPO:handbook,graGraBook} and
single~\cite{SPO:handbook} pushout (DPO and SPO) which have developed
interesting analysis techniques, for example to check sequential and
parallel independence between pairs of
rules~\cite{graGraBook,handbook} or the calculation of critical
pairs~\cite{Heckel,Lambers}.

Frequently, graph transformation rules are equipped with {\em
  application conditions} (ACs)~\cite{AC:Ehrig,graGraBook,HeckelW95},
stating extra (in addition to the left hand side) positive and
negative conditions that the host graph should satisfy for the rule to
be applicable.  The algebraic approach has proposed a kind of ACs with
predefined diagrams (i.e. graphs and morphisms making the condition)
and quantifiers regarding the existence or not of matchings of the
different graphs of the constraint in the host
graph~\cite{AC:Ehrig,graGraBook}. Most analysis techniques for plain
rules (without ACs) have to be adapted then for rules with ACs (see
e.g.~\cite{Lambers} for critical pairs with negative ACs). Moreover,
different adaptations may be needed for different kinds of ACs. Thus,
a uniform approach to analyze rules with arbitrary ACs would be very
useful.

In previous works~\cite{JuanPP_1,JuanPP_2,JuanPP_4,MGGBook} we
developed a framework (Matrix Graph Grammars, MGGs) for the
transformation of simple digraphs. Simple digraphs and their
transformation rules can be represented using Boolean matrices and
vectors. Thus, the rewriting can be expressed using Boolean operators
only. One important point is that, as a difference from other
approaches, we explicitly represent the rule dynamics (addition and
deletion of elements) instead of only the static parts (rule pre and
postconditions). This point of view enables new analysis techniques,
such as for example checking independence of a sequence of arbitrary
length and a permutation of it, or to obtain the smallest graph able
to fire a sequence. On the theoretical side, our formalization of
graph transformation introduces concepts from many branches of
mathematics like Boolean algebra, group theory, functional analysis,
tensor algebra and logics~\cite{MGGBook, MGGCombinatorics, MGGmodel}.
This wealth of available mathematical results opens the door to new
analysis methods not developed so far, like sequential independence
and explicit parallelism not limited to pairs of sequences,
applicability, graph congruence and reachability. On the practical
side, the implementations of our analysis techniques, being based on
Boolean algebra manipulations, are expected to have a good
performance.

In MGGs we do not only consider the elements that must be present in
order to apply a production (left hand side, LHS, also known as
\emph{certainty part}) but also those elements that potentially
prevent its application (also known as \emph{nihil} or
\emph{nihilation part}). Refer to~\cite{MGGfundamenta} in which,
besides this, application conditions and graph constraints are studied
for the MGG approach. The present contribution is a continuation
of~\cite{MGGfundamenta} where a comparison with related work can also be
found. We shall tackle pre and postconditions, their transformation,
the sequential version of these results and multidigraph rewriting.

\noindent {\bf Paper organization}. Section~\ref{sec:MGGs} gives an
overview of Matrix Graph Grammars. Section~\ref{sec:previous} revises
application conditions as studied
in~\cite{MGGfundamenta}. Postconditions and their equivalence to
certain sequences are addressed in
Sec.~\ref{sec:postconditions}. Section~\ref{sec:movingConditions}
tackles the transformation of preconditions into postconditions. The
converse, more natural from a practical point of view, is also
addressed. The transformation of restrictions is generalized in
Sec.~\ref{sec:delocalization} in which \emph{delocalization} -- how to
move application conditions from one production to another inside the
same sequence -- is also studied together with \emph{variable
  nodes}. As an application of restrictions to MGGs,
Sec.~\ref{sec:fromSimpleDigraphsToMultidigraphs} shows how to make MGG
deal with multidigraphs instead of just simple digraphs without major
modifications to the theory. The paper ends in
Sec.~\ref{sec:conclusions} with some conclusions, further research
remarks and acknowledgements.

\section{Matrix Graph Grammars Overview}
\label{sec:MGGs}

\begin{figure}[htbp]
  \centering
  \includegraphics[scale = 0.4]{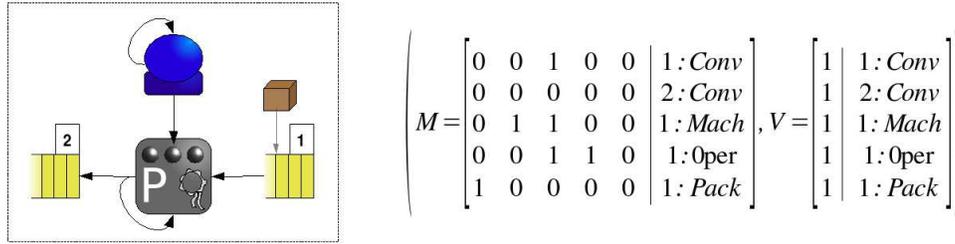}
  \caption{Simple Digraph Example (left). Matrix Representation
    (right)}
  \label{fig:example_graph}
\end{figure}

We work with \textbf{simple digraphs} which we represent as $G =
(M, V)$, where $M$ is a Boolean matrix for edges (the graph {\em
  adjacency} matrix) and $V$ a Boolean vector for vertices or
nodes.\footnote{The vector for nodes is necessary because in MGG nodes
  can be added and deleted, and thus we mark the existing nodes with a
  $1$ in the corresponding position of the vector.} The left of
Fig.~\ref{fig:example_graph} shows a graph representing a production
system made up of a machine (controlled by an operator) which consumes
and produces pieces through conveyors. Self loops in operators and
machines indicate that they are busy.

Well-formedness of graphs (i.e. absence of dangling edges) can be
checked by verifying the identity $\left\| \left( M \vee M^t \right)
  \odot \overline{V}\right\| _1 = 0$, where $\odot$ is the Boolean
matrix product,\footnote{The Boolean matrix product is like the regular
  matrix product, but with \textbf{and} and \textbf{or} instead of
  multiplication and addition.} $M^t$ is the transpose of the matrix
$M$, $\overline{V}$ is the negation of the nodes vector $V$, and $\|
\cdot \|_1$ is an operation (a norm, actually) that results in the
{\bf or} of all the components of the vector. We call this property
\textbf{compatibility} (refer to~\cite{JuanPP_1}). Note that $M \odot
\overline{V}$ results in a vector that contains a 1 in position $i$
when there is an outgoing edge from node $i$ to a non-existing node. A
similar expression with the transpose of $M$ is used to check for
incoming edges.

A \textbf{type} is assigned to each node in $G=(M,V)$ by a function
from the set of nodes $|V|$ to a set of types $T$,
$\abb{\lambda}{|V|}{T}$. Sets will be represented by $|\cdot|$. In
Fig.~\ref{fig:example_graph} types are represented as an extra column
in the matrices, where the numbers before the colon distinguish
elements of the same type. It is just a visual aid. For edges we use
the types of their source and target nodes. A \textbf{typed simple
  digraph} is $G_T=(G, \lambda)$. From now on we shall assume typed
graphs and shall drop the $T$ subindex.

A \textbf{production} or grammar rule $p:L \rightarrow R$ is a
morphism of typed simple digraphs, which is defined as a mapping that
transforms $L$ in $R$ with the restriction that the type of the image
must be equal to the type of the source element.\footnote{We shall
  come back to this topic in Sec.~\ref{sec:delocalization}.} More
explicitly, $\abb{f=(f_V, f_E)}{G_1}{G_2}$ being $f_V$ and $f_E$
partial injective mappings $\abb{f_V}{|V_1|}{|V_2|}$,
$\abb{f_E}{|M_1|}{|M_2|}$ such that $\forall v \in Dom(f_V), \:
\lambda_1(v) = \lambda_2(f_V(v))$ and $\forall e=(n, m) \in Dom(f_E),
\: f_E(e) = f_E(n, m)=(f_V(n), f_V(m))$, where $Dom$ stands for
domain, $E$ for edges and $V$ for vertices.

A production $p:L \rightarrow R$ is \textbf{statically represented} as
$p = (L, R) = \left( \left( L^E, L^V, \lambda^L \right), \left( R^E,
  R^V, \lambda^R \right) \right)$. The matrices and vectors of these graphs are arranged so
that the elements identified by morphism $p$ match (this is called
completion, see below). Alternatively, a production adds and deletes
nodes and edges, therefore they can be \textbf{dynamically represented}
by encoding the rule's LHS together with matrices and vectors
representing the addition and deletion of edges and nodes:\footnote{We
  call such matrices and vectors $e$ for ``erase'' and $r$ for
  ``restock''.} $p = (L, e, r) = \left( \left( L^E, L^V, \lambda^L
  \right), e^E, r^E, e^V, r^V, \lambda^r \right)$, where $\lambda^r$
contains the types of the new nodes, $e^E$ and $e^V$ are the deletion
Boolean matrix and vector, $r^E$ and $r^V$ are the addition Boolean
matrix and vector. They have a 1 in the position where the element is
to be deleted or added, respectively. The output of rule $p$ is
calculated by the Boolean formula $R = p(L) = r \vee \overline{e} \,
L$, which applies both to nodes and edges.\footnote{The and symbol
  $\wedge$ is usually omitted in formulae, so $R = p(L) = r \vee
  \overline{e} \wedge L$ with precedence of $\wedge$ over $\vee$.}

\begin{figure}[htbp]
  \centering
  \includegraphics[scale = 0.22]{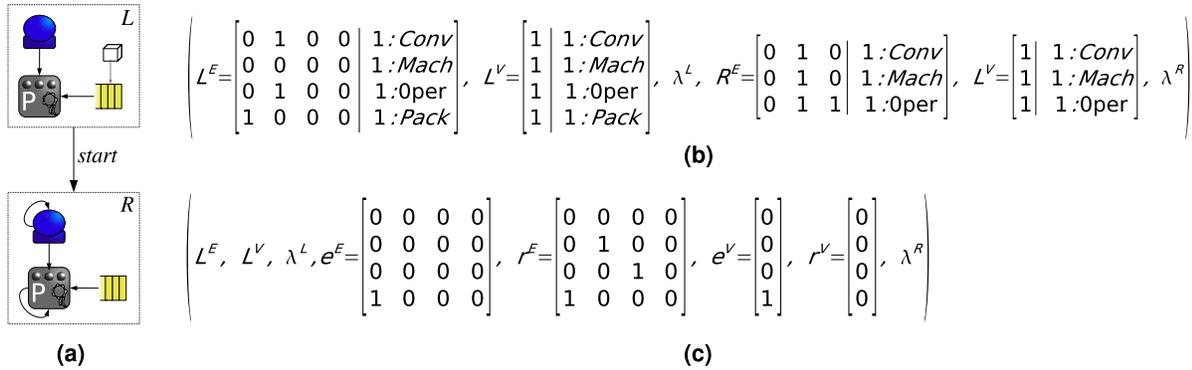}
  \caption{(a) Rule Example. (b) Static Formulation. (c) Dynamic
    Formulation}
  \label{fig:example_rule}
\end{figure}

\noindent \textbf{Example}.$\square$Figure~\ref{fig:example_rule}
shows a rule and its associated matrices. The rule models the
consumption of a piece (Pack) by a machine (Mach) input via the
conveyor (Conv). There is an operator (Oper) managing the
machine. Compatibility of the resulting graph must be ensured, thus
the rule cannot be applied if the machine is already busy, as it would
end up with two self loops which is not allowed in a simple digraph.
This restriction of simple digraphs can be useful in this kind of
situations and acts like a built-in negative application
condition. Later we will see that the \emph{nihilation matrix} takes
care of this restriction. \proofend

In order to operate with the matrix representation of graphs of
different sizes, an operation called \textbf{completion} adds extra
rows and columns with zeros to matrices and vectors, and rearranges
rows and columns so that the identified edges and nodes of the two
graphs match.  For example, in Fig.~\ref{fig:example_rule}, if we need
to operate $L^E$ and $R^E$, completion adds a fourth $0$-row and
fourth $0$-column to $R^E$. No further modification is needed because
the rest of the elements have the right types and are placed
properly.\footnote{In the present contribution we shall assume that
  completion is being performed somehow. This is closely related to
non-determinism. The reader is referred to~\cite{MGGmodel} for further
details.}

With the purpose of considering the elements in the host graph that
disable a rule application, we extend the notation for rules with a
new simple digraph $K$, which specifies the two kinds of forbidden
edges: Those incident to nodes which are going to be erased and any
edge added by the rule (which cannot be added twice, since we are
dealing with simple digraphs). $K$ has non-zero elements in positions
corresponding to newly added edges, and to non-deleted edges incident
to deleted nodes. Matrices are derived in the following order: $\left(
  L, R \right) \mapsto \left( e, r \right) \mapsto K$.  Thus, a rule
is \emph{statically} determined by its LHS and RHS $p = \left( L, R
\right)$, from which it is possible to give a \emph{dynamic}
definition $p = \left(L, e, r \right)$, with $e = L\overline{R}$ and
$r = R\overline{L}$, to end up with a full specification including its
\emph{environmental} behavior $p = \left(L, K, e, r \right)$.  No
extra effort is needed from the grammar designer because $K$ can be
automatically calculated: $K = p ( \overline{D} )$, with $D =
\overline{e^V} \otimes \overline{e^V}^{\,t}$.\footnote{Symbol
  $\otimes$ denotes the tensor or Kronecker product, which sums up the
  covariant and contravariant parts and multiplies every element of
  the first vector by the whole second vector.} The evolution of the
nihilation matrix (what elements can not appear in the RHS) -- call it
$Q$ -- is given by the inverse of the production: $(R,Q) = \left(
  p(L), p^{-1}(K) \right) = \left( r \vee \overline{e} L, e \vee
  \overline{r} K \right)$. See~\cite{MGGfundamenta} for more details.

Inspired by the Dirac or bra-ket notation~\cite{braket} we split the
static part (initial state, $L$) from the dynamics (element addition
and deletion, $p$): $R = p(L) = \left \langle L, p \right
\rangle$. The \emph{ket} operators (those to the right side of the
bra-ket) can be moved to the \emph{bra} (left hand side) by using
their adjoints.

\textbf{Matching} is the operation of identifying the LHS of a rule
inside a host graph. Given a rule $p:L \rightarrow R$ and a simple
digraph $G$, any total injective\footnote{MGG considers only injective
  matches.} morphism $m:L \rightarrow G$ is a match for $p$ in $G$,
thus it is one of the ways of {\em completing} $L$ in $G$. Besides, we
shall consider the elements that must not be present.

Given the grammar rule $p:L \rightarrow R$ and the graph $G = (G^E,
G^V)$, $d = \left( p, m \right)$ is called a \textbf{direct
  derivation} with $m = \left( m_L, m_K \right)$ and result $H = p^*
\left( G \right)$ if the following conditions are satisfied:
\begin{enumerate}
\item There exist total injective morphisms $m_L : L \rightarrow G$
  and $m_K : K \rightarrow \overline{G}$ with $m_L(n) = m_K(n)$,
  $\forall n \in L^V$.
\item The match $m_L$ induces a completion of $L$ in $G$. Matrices
  $e$ and $r$ are then completed in the same way to yield $e^*$ and
  $r^*$. The output graph is calculated as $H = p^*(G) = r^* \vee
  \overline {e^*} G$.
\end{enumerate}

The negation when applied to graphs alone (not specifying the nodes)
-- e.g. $\overline{G}$ in the first condition above -- will be carried
out just on edges. Notice that in particular the first condition above
guarantees that $L$ and $K$ will be applied to the same nodes in the
host graph $G$.

In direct derivations \emph{dangling edges} can occur because the
nihilation matrix only considers edges incident to nodes appearing in
the rule's LHS and not in the whole host graph. In MGG an operator
$T_\varepsilon$ takes care of dangling edges which are deleted by
adding a preproduction (known as $\varepsilon-$production) before the
original rule. Refer to~\cite{JuanPP_1, JuanPP_2}. Thus, rule $p$ is
transformed into the sequence $p;p_{\varepsilon}$, where
$p_\varepsilon$ deletes the dangling edges and $p$ remains
unaltered.

There are occasions in which two or more productions should be matched
to the same nodes. This is achieved with the \textbf{marking operator}
$T_\mu$ introduced in Chap.~6 in~\cite{MGGBook}. A grammar rule and
its associated $\varepsilon$-production is one example and we shall
find more in future sections.

In~\cite{JuanPP_1,JuanPP_2,JuanPP_4,MGGBook} some analysis techniques
for MGGs have been developed which we shall skim through. One
important feature of MGG is that sequences of rules can be analyzed
independently to some extent of any host graph. A rule
\textbf{sequence} is represented by $s_n = p_n;\ldots; p_1$ where
application is from right to left, i.e. $p_1$ is applied first. For
its analysis, the sequence is completed by identifying the nodes
across rules which are assumed to be mapped to the same node in the
host graph.

Once the sequence is completed, sequence
\textbf{coherence}~\cite{JuanPP_1, MGGBook, MGGCombinatorics}
allows us to know if, for the given identification, the sequence is
potentially applicable, i.e. if no rule disturbs the application of
those following it. The formula for coherence results in a matrix and
a vector (which can be interpreted as a graph) with the problematic
elements. If the sequence is coherent, both should be zero; if not,
they contain the problematic elements. A coherent sequence is
\textbf{compatible} if its application produces a simple digraph. That
is, no dangling edges are produced in intermediate steps.

Given a completed sequence, the \textbf{minimal initial digraph} (MID)
is the smallest graph that permits the application of such
sequence. Conversely, the \textbf{negative initial digraph} (NID)
contains all elements that should not be present in the host graph for
the sequence to be applicable. In this way, the NID is a graph that
should be found in $\overline G$ for the sequence to be applicable
(i.e. none of its edges can be found in $G$). See Sec.~6
in~\cite{MGGCombinatorics} or Chaps.~5 and~6 in~\cite{MGGBook}.

Other concepts we developed aim at checking \textbf{sequential
  independence} (same result) between a sequence and a permutation of
it. \textbf{G-Congruence} detects if two sequences, one permutation of
the other, have the same MID and NID. It returns two matrices and two
vectors, representing two graphs which are the differences between the
MIDs and NIDs of each sequence, respectively. Thus if zero, the
sequences have the same MID and NID. Two coherent and compatible
completed sequences that are G-congruent are sequentially
independent. See Sec.~7 in~\cite{MGGCombinatorics} or Chap.~7
in~\cite{MGGBook}.

\section{Previous Work on Application Conditions in MGG}
\label{sec:previous}

In this section we shall brush up on application conditions (ACs) as
introduced for MGG in~\cite{MGGfundamenta} with non-fixed diagrams and
quantifiers. For the quantification, a full-fledged monadic second
order logic\footnote{MSOL, see e.g. \cite{Courcelle}.} formula is
used. One of the contributions in~\cite{MGGfundamenta} is that a
rule with an AC can be transformed into (sequences of) plain rules by
adding the positive information to the left hand side of the
production and the negative to the nihilation matrix.

A \textbf{diagram} $\mathfrak{d}$ is a set of simple digraphs $\{ A^i
\}_{i \in I}$ and a set of partial injective morphisms $\{ d_k \}_{k
  \in K}$ with $d_k: A^i \rightarrow A^j$. The diagram $\mathfrak{d}$
is well defined if every cycle of morphisms commute. $GC = \big(
\mathfrak{d} = ( \{A^i \}_{i \in I}, \{ d_j \}_{j \in J} ),
\mathfrak{f} \big)$ is a \textbf{graph constraint} where
$\mathfrak{d}$ is a well defined diagram and $\mathfrak{f}$ a sentence
with variables in $\{ A^i \}_{i \in I}$ and predicates $P$ and
$Q$. See eqs.~(\ref{eq:P}) and~(\ref{eq:Q}). Formulae are restricted
to have no free variables except for the default second argument of
predicates $P$ and $Q$, which is the host graph $G$ in which we
evaluate the GC. GC formulae are made up of expressions about graph
inclusions. The predicates $P$ and $Q$ are given by:
\begin{align}
  \label{eq:P}
  P(X^1, X^2) &= \forall m [F(m, X^1) \Rightarrow F(m, X^2)] \\
  \label{eq:Q}
  Q(X^1, X^2) &= \exists e [F(e, X^1) \wedge F(e, X^2)],
\end{align}
where predicate $F(m, X)$ states that element $m$ (a node or an edge)
is in graph $X$.  Predicate $P(X^1, X^2)$ means that graph $X^1$ is
included in $X^2$. Predicate $Q(X^1, X^2)$ asserts that there is a
partial morphism between $X^1$ and $X^2$, which is defined on at least
one edge ($e$ ranges over all edges). The notation (syntax) will be
simplified by making the host graph $G$ the default second argument
for predicates $P$ and $Q$.  Besides, it will be assumed that by
default total morphisms are demanded: Unless otherwise stated
predicate $P$ is assumed. We take the convention that negations in
abbreviations apply to the predicate (e.g. $\exists A [\overline{A}]
\equiv \exists A [\overline{P}(A, G)]$) and not the negation of the
graph's adjacency matrix.

\begin{figure}[htbp]
  \centering
  \includegraphics[scale = 0.35]{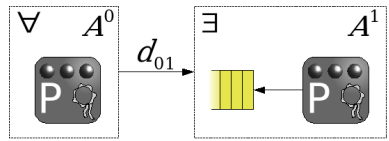}
  \caption{Diagram Example}
  \label{fig:DiagExample}
\end{figure}

\noindent {\bf Example}.$\square$The GC in Fig.~\ref{fig:DiagExample}
is satisfied if for every $A^0$ in $G$ it is possible to find a
related $A^1$ in $G$, i.e. its associated formula is $\forall A^0
\exists A^1 \left[ A^0 \Rightarrow A^1\right]$, equivalent by
definition to $\forall A^0 \exists A^1 \left[ P\!\!\left(A^0, G
  \right) \Rightarrow P\!\! \left(A^1, G\right) \right]$. Nodes and
edges in $A^0$ and $A^1$ are related through morphism $d_{01}$ in
which the image of the machine in $A^0$ is the machine in $A^1$. To
enhance readability, each graph in the diagram has been marked with
the quantifier given in the formula. The GC in
Fig.~\ref{fig:DiagExample} expresses that each machine should have an
output conveyor. \proofend

Given the rule $p:L \rightarrow R$ with nihilation matrix $K$, an
\textbf{application condition} AC (over the free variable $G$) is a
GC satisfying:
\begin{enumerate}
\item $\exists ! i,j$ such that $A^i = L$ and $A^j = K$.
\item $\exists ! k$ such that $A^k = G$ is the only free variable.
\item $\mathfrak{f}$ must demand the existence of $L$ in $G$ and the
  existence of $K$ in $\overline{G}$.
\end{enumerate}

For simplicity, we usually do not explicitly show the condition 3 in
the formulae of ACs, nor the nihilation matrix $K$ in the diagram
which are existentially quantified before any other graph of the
AC. Notice that the rule's LHS and its nihilation matrix can be
interpreted as the minimal AC a rule can have. For technical reasons
addressed in Sec.~\ref{sec:movingConditions} (related to converting
pre into postconditions) we assume that morphisms $d_i$ in the diagram
do not have codomain $L$ or $K$. This is easily solved as we may
always use their inverses due to $d_i$'s injectiveness.

It is possible to embed arbitrary ACs into rules by including the
positive and negative conditions in $L$ and $K$, respectively.
Intuitively: ``MGG + AC = MGG'' and ``MGG + GC =
MGG''. In~\cite{MGGfundamenta} two basic operations are
introduced:~\textbf{closure} -- $\widecheck{T}$ -- that transforms
universal into existential quantifiers, and \textbf{decomposition} --
$\widehat{T}$ -- that transforms partial morphisms into total
morphisms. Notice that a match is an existentially quantified total
morphism. It is proved in~\cite{MGGfundamenta} that any AC can be
embedded into its corresponding direct derivation. This is achieved by
transforming the AC into some sequences of productions. There are four
basic types of ACs/GCs. Let $GC = \left( \mathfrak{d}, \mathfrak{f}
\right)$ be a graph constraint with diagram $\mathfrak{d} = \{ A \}$
and consider the associated production $p:L \to R$. The case
$\mathfrak{f} = \exists A[A]$ is just the matching of $A$ in the host
graph $G$. It is equivalent to the sequence $p ; id_A$, where $id_A$
has $A$ as LHS and RHS, so it simply demands its existence in $G$. We
introduce the operator $T_A$ that replaces $p$ by $p;id_A$ and leaves
the diagram and the formula unaltered. If the formula $\mathfrak{f} =
\forall A [A]$ is considered, we can reduce it to a sequence of
matchings via the closure operator $\widecheck{T}_A$ whose result is:
\begin{align}
  \label{eq:closure}
  \mathfrak{d} & \longmapsto \mathfrak{d}' = \left(\{ A^1, \ldots,
    A^n \}, d_{ij}:A^i \rightarrow A^j \right) \nonumber \\
  \mathfrak{f} & \longmapsto \mathfrak{f}' = \exists A^1 \ldots
  \exists A^n \left[ \bigwedge_{i=1}^n A^i \right],
\end{align} 
with $A^i \cong A$, $d_{ij} \not \in iso(A^i, A^j)$,\footnote{$iso(A,
  G) = \{ \abb{f}{A}{G} \; | \; f$ is an isomorphism$\}$.}
$\widecheck{T}_A \left( AC \right) = AC' = \left( \mathfrak{d}',
  \mathfrak{f}' \right)$ and $n=|par^{max}(A,
G)|$.\footnote{$par^{max}(A, G) = \{ \abb{f}{A}{G} \; | \; f$ is a
  maximal non-empty partial morphism with $Dom(f)^V=A^V \}$.} This is
equivalent to the sequence $p;id_{A^n};\ldots;id_{A^1}$. If the
application condition has formula $\mathfrak{f} = \exists A [Q(A)]$,
we can proceed by defining the composition operator $\widehat{T}_A$
with action:
\begin{align}
  \label{eq:decomp}
  \mathfrak{d} & \longmapsto \mathfrak{d}' = \left(\{ A^1, \ldots,
    A^n \}, d_{ij}:A^i \rightarrow A^j \right) \nonumber \\
  \mathfrak{f} & \longmapsto \mathfrak{f}' = \exists A^1 \ldots
  \exists A^n \left[ \bigvee_{i=1}^n A^i \right],
\end{align}
where $A^i$ contains a single edge of $A$ and $n$ is the number of
edges of $A$. This is equivalent to the set of sequences $\left\{
  p;id_{A^i} \right\}, i \in \{1, \ldots, n\}$.

Less evident are formulas of the form $\mathfrak{f} = \nexists A [A] =
\forall A [ \overline{A} ]$. Fortunately, operators $\widecheck{T}$
and $\widehat{T}$ commute when composed so we can get along with the
operator $\widetilde{T}_A = \widehat{T}_A \circ \widecheck{T}_A =
\widecheck{T}_A \circ \widehat{T}_A$. The image of $\widetilde{T}$ on
such ACs are given by:
\begin{align}
  \label{eq:nac}
  \mathfrak{d} & \longmapsto \mathfrak{d}' = \left(\{ A^{11}, \ldots,
    A^{mn} \}, d_{ij}:A^i \rightarrow A^j \right) \nonumber \\
  \mathfrak{f} & \longmapsto \mathfrak{f}' = \exists A^{11} \ldots
  \exists A^{mn} \left[ \bigwedge^{m}_{i=1}\bigvee_{j=1}^n P \left(
      A^{ij}, \overline{G} \right) \right].
\end{align}

An AC is said to be \textbf{coherent} if it is not a contradiction
(false in all scenarios), \textbf{compatible} if, together with the
rule's actions, produces a simple digraph, and \textbf{consistent} if
$\exists G$ host graph such that $G \models AC$\footnote{We shall say
  that the host graph $G$ satisfies $\exists A[A]$, written $G \models
  \exists A[A]$, if and only if $\exists f \in par^{max}(A, G) \: [ f
  \in tot(A, G)]$, being $tot(A, G) = \{ f:A \to G \; | \; f$ is a
  total morphism$\} \subseteq par^{max}(A, G)$. Also, $G$ satisfies
  $\forall A[A]$, written\phantom{i} $G \models \forall A[A]$, if and
  only if $\forall f \in par^{max}(A, G) \: [f \in tot(A,
  G)]$. Usually we shall abuse of the notation and write $G \models
  GC$ instead. For more details, please refer
  to~\cite{MGGfundamenta}.} to which the production is applicable. As
ACs can be transformed into equivalent (sets of) sequences, it is
proved in~\cite{MGGfundamenta} that coherence and compatibility of an
AC is equivalent to coherence and compatibility of the associated (set
of) sequence(s), respectively. Also, an AC is consistent if and only
if its equivalent (set of) sequence(s) is applicable. Besides, all
results and analysis techniques developed for MGG can be applied to
sequences with ACs. Some examples follow:
\begin{itemize}
\item As a sequence is applicable if and only if it is coherent and
  compatible (see Sec~6.4 in~\cite{MGGBook}) then an AC is consistent
  if and only if it is coherent and compatible.
\item Sequential independence allows us to delay or advance the
  constraints inside a sequence. As long as the productions do not
  modify the elements of the constraints, this is transformation of
  preconditions into postconditions. More on
  Sec.~\ref{sec:movingConditions}.
\item Initial digraph calculation solves the problem of finding a host
  graph that satisfies a given AC/GC. There are some limitations,
  though. For example it is necessary to limit the maximum number of
  nodes when dealing with universal quantifiers. This has no impact in
  some cases, for example when non-uniform MGG submodels are
  considered (see \emph{nodeless MGG} in~\cite{MGGmodel}).
\item \emph{Graph congruence} characterizes sequences with the same
  initial digraph. Therefore it can be used to study when two
  GCs/ACs are equivalent for all morphisms or for some of them.
\end{itemize}

Summarizing, there are two basic results
in~\cite{MGGfundamenta}. First, it is always possible to embed an
application condition into the LHS of the production or
derivation. The left hand side $L$ of a production receives elements
that must be found -- $P(A,G)$ -- and $K$ those whose presence is
forbidden -- $\overline{P}(A,G)$ --. Second, it is always possible to
find a sequence or a set of sequences of plain productions whose
behavior is equivalent to that of the production plus the application
condition.

\section{Postconditions}
\label{sec:postconditions}

In this section we shall introduce postconditions and state some basic
facts about them analogous to those for preconditions. We shall
enlarge the notation by appending a left arrow on top of the
conditions to indicate that they are preconditions and an upper
right arrow for postconditions.  Examples are
$\stackrel{\leftarrow}{A}$ for a precondition and
$\stackrel{\rightarrow}{A}$ for a postcondition.  If it is clear
from the context, arrows will be omitted.

\begin{definition}[Precondition and Postcondition]
  \label{def:prePostCondition}
  An application condition set on the LHS of a production is known as
  a \emph{precondition}. If it is set on the RHS then it is known as a
  \emph{postcondition}.
\end{definition}

Operators $T_{\stackrel{\rightarrow}{A}},
\widehat{T}_{\stackrel{\rightarrow}{A}},
\widecheck{T}_{\stackrel{\rightarrow}{A}}$ and
$\widetilde{T}_{\stackrel{\rightarrow}{A}}$ are defined similarly for
postconditions. The following proposition establishes an
equivalence between the basic formulae (match, decomposition, closure
and negative application condition) and certain sequences of
productions.

\begin{proposition}
  \label{prop:postConds}
  Let $\stackrel{\rightarrow}{A} \, = \left( \mathfrak{f},
    \mathfrak{d} \right) = \big( \mathfrak{f}, \left(\{A\}, d:R
    \rightarrow A \right)\!\big)$ be a postcondition. Then we can
  obtain a set of equivalent sequences to given basic formulae as
  follows:
  \begin{align}
    \label{eq:postMatch}
    \textrm{(Match)} \quad \mathfrak{f} &= \exists A [A] \quad
    \longmapsto \quad T_A\left(p\right) = id_A ; p \\
    \label{eq:closure}
    \textrm{(Closure)} \quad
    \mathfrak{f} &= \nexists A [\overline{A}] \quad
    \longmapsto \quad \widecheck{T}_A\left(p\right) = id_{A^1} ;
    \ldots ; id_{A^m}; p \\
    \label{eq:decomp}
    \textrm{(Decomposition)} \quad 
    \mathfrak{f} &= \exists A [\overline{A}] \quad \longmapsto
    \quad \widehat{T}_A\left(p\right) = \left\{ \overline{id}_{A_i} ;
      p \right\}_{i = 1, \ldots, n} \\
    \label{eq:postNAC} \textrm{(NAC)} \quad \mathfrak{f}
    &= \nexists A [A] \quad \longmapsto \quad 
    \widetilde{T}_A \left( p \right) = \left\{
      \overline{id}_{A^1_{i_1}} ; \ldots ;
      \overline{id}_{A^m_{i_m}}; p \right\}_{i_j \in \{1, \ldots, n\},
    j \in \{1, \ldots, m\}}
  \end{align}
  where $m$ is the number of potential matches of $A$ in the image of
  the host graph, $n$ is the number of edges in $A$ and
  $\overline{id}_A$ asks for the existence of $A$ in the complement of
  the image of the host graph.
\end{proposition}

\noindent \emph{Proof} \\*
$\square$ For the first case (match), the AC states that an additional
graph $A$ has to be found in the image of the host graph. This is
easily achieved by applying $id_A$ to the image of $L$, i.e. by
considering $id_A \left( p(L) \right) = \left( id_A \circ p \right)
(L)$. The elements in $A$ are related to those in $R$ according to the
identifications in a morphism $d$ that has to be given in the diagram
of the postcondition. In the four cases considered in the proposition
we can move from composition to concatenation by means of the marking
operator $T_\mu$. Recall that $T_\mu$ guarantees that the
identifications in $d$ are preserved.

The second case (closure) is very similar. We have to verify all
potential appearances of $A$ in the image of the host graph because
$\nexists A [\overline{A}] = \forall A [A]$. We proceed as in the
first case but this time with a finite number of compositions: $\left(
  id_{A^1} \circ \ldots \circ id_{A^m} \circ p \right) (L)$.

For decomposition, $A$ is not found in the host graph if for some
matching there is at least one missing edge. It is thus similar to
matching but for a single edge. The way to proceed is to consider the
set of sequences that appear in eq.~(\ref{eq:decomp}). Negative
application conditions (NACs) are the composition of
eqs.~(\ref{eq:closure}) and~(\ref{eq:decomp}). \proofend

One of the main points of the techniques available for preconditions
is to analyze rules with ACs by translating them into sequences of
flat rules, and then analyzing the sequences of flat rules instead.

\begin{theorem}
  \label{th:reductionPost}
  Any well-defined postcondition can be reduced to the study of
  the corresponding set of sequences.
\end{theorem}

\noindent \emph{Proof} \\*
$\square$The proof follows that of Th.~4.1 in~\cite{MGGfundamenta} and
is included here for completeness sake. Let the depth of a graph for a
fixed node $n_0$ be the maximum over the shortest path (to avoid
cycles) starting in any node different from $n_0$ and ending in
$n_0$. The depth of a graph is the maximum depth for all its
nodes. Notice that the depth is $1$ if and only if $A^i, \forall i$ in
the diagram are unrelated. We shall apply induction on the depth of
the AC.

A diagram $\mathfrak{d}$ is a graph where nodes are digraphs $A^i$ and
edges are morphisms $d_{ij}$. There are $16$ possibilities for depth
$1$ in a AC made up of a single element $A$, summarized in
Table~\ref{tab:possibilitiesSingleCase}.

\begin{table*}[hbtp]
  \centering
  \begin{tabular}{|rl|rl||rl|rl|}
    \hline
    \begin{Large}\phantom{I}\end{Large}(1*) & $\exists A [ A ]$ & (5*)
    & $\slash\!\!\forall A [\overline{A}]$ & (9*) & $\exists A
    [\overline{Q}(A)]$ & (13*) & $\slash\!\!\forall A [Q(A)]$\\
    \hline
    \begin{Large}\phantom{I}\end{Large}(2*) & $\exists A
    [\overline{A}]$ & (6*) & $\slash\!\!\forall A [A]$ & (10*) &
    $\exists A [Q(A)]$ & (14*) & $\slash\!\!\forall A
    [\overline{Q}(A)]$\\ 
    \hline
    \begin{Large}\phantom{I}\end{Large}(3*) & $\nexists A
    [\overline{A}]$ & (7*) & $\forall A [A]$ & (11*) & $\nexists A
    [Q(A)]$ & (15*) & $\forall A [\overline{Q}(A)]$\\
    \hline
    \begin{Large}\phantom{I}\end{Large}(4*) & $\nexists A [ A ]$ &
    (8*) & $\forall A [\overline{A}]$ & (12*) & $\nexists A
    [\overline{Q}(A)]$ & (16*) & $\forall A [Q(A)]$\\
    \hline
  \end{tabular}
  \caption{All Possible Diagrams for a Single Element}
  \label{tab:possibilitiesSingleCase}
\end{table*}

Elements in the same row for each pair of columns are related using
equalities $\nexists A[A] = \forall A[\overline{A}]$ and
$\slash\!\!\forall A[A] = \exists A[\overline{A}]$, so it is possible
to reduce the study to cases (1*) -- (4*) and (9*) --
(12*). Identities $\overline{Q}(A) = P(A, \overline{G})$ and $Q(A) =
\overline{P}(A, \overline{G})$ reduce (9*) -- (12*) to formulae (1*)
-- (4*):
\begin{align}
  \exists A [\overline{Q}(A)] = \exists A\left[P(A,
    \overline{G})\right] &, \; \exists A [Q(A)] = \exists
  A\left[\overline{P}(A, \overline{G})\right] \nonumber \\
  \nexists A [Q(A)] = \nexists A\left[\overline{P}(A,
    \overline{G})\right] &, \; \nexists A [\overline{Q}(A)] = \nexists
  A\left[P(A, \overline{G})\right]. \nonumber
\end{align}

Proposition~\ref{prop:postConds} considers the four basic cases which
correspond to (1*) -- (4*) in Table~\ref{tab:possibilitiesSingleCase},
showing that in fact they can all be reduced to matchings in the image
of the host graph, i.e. to~(1*) in
Table~\ref{tab:possibilitiesSingleCase}, verifying the theorem.

Now we move on to the induction step which considers combinations of
quantifiers. Well-definedness guarantees independence with respect to
the order in which elements $A^i$ in the postcondition are
selected. When there is a universal quantifier $\forall A$, according
to eq.~(\ref{eq:closure}), elements of $A$ are replicated as many
times as potential instances of $A$ can be found in the host graph.
In order to continue the procedure we have to clone the rest of the
diagram for each replica of $A$, except those graphs which are
existentially quantified before $A$ in the formula. That is, if we
have a formula $\exists B \forall A \exists C$ when performing the
closure of $A$, we have to replicate $C$ as many times as $A$, but not
$B$.  Moreover $B$ has to be connected to each replica of $A$,
preserving the identifications of the morphism $B \rightarrow A$.
More in detail: When closure is applied to $A$, we iterate on all
graphs $B_j$ in the diagram. There are three possibilities:

\begin{itemize}
\item If $B_j$ is existentially quantified after $A$ -- $\forall A ...
  \exists B_j$ -- then it is replicated as many times as $A$.
  Appropriate morphisms are created between each $A^i$ and $B^i_j$ if
  a morphism $\abb{d}{A}{B}$ existed. The new morphisms identify
  elements in $A^i$ and $B^i_j$ according to $d$.  This permits finding
  different matches of $B_j$ for each $A^i$, some of which can be
  equal.\footnote{If for example there are three instances of $A$ in
    the image of the host graph but only one of $B_j$, then the three
    replicas of $B$ are matched to the same part of $p(G)$.}
\item If $B_j$ is existentially quantified before $A$ -- $\exists B_j
  ... \forall A$ -- then it is not replicated, but just connected to
  each replica of $A$ if necessary. This ensures that a unique $B_j$
  has to be found for each $A^i$. Moreover, the replication of $A$ has
  to preserve the shape of the original diagram. That is, if there is
  a morphism $\abb{d}{B}{A}$ then each $\abb{d_i}{B}{A^i}$ has to
  preserve the identifications of $d$ (this means that we take only
  those $A^i$ which preserve the structure of the diagram).
\item If $B_j$ is universally quantified (no matter if it is
  quantified before or after $A$), again it is replicated as many
  times as $A$. Afterwards, $B_j$ will itself need to be replicated
  due to its universality. The order in which these replications are
  performed is not relevant as $\forall A \forall B_j = \forall B_j
  \forall A$. \proofend
\end{itemize}

Previous theorem and the corollaries that follow heavily depend on the
host graph and its image (through matching) so analysis techniques
developed so far in MGG which are independent of the host graphs can
not be applied. The ``problem'' is the universal quantifier. We can
consider the initial digraph and dispose to some extent of the host
graph and its image. This is related to the fact
(Sec.~\ref{sec:movingConditions}) that it is possible to transform
postconditions into equivalent preconditions.

Two applications of Th.~\ref{th:reductionPost} are the
following corollaries that characterize coherence, compatibility and
consistency of postconditions.

\begin{corollary}
  \label{cor:equivPostAC_seqs}
  A postcondition is coherent if and only if its associated (set of)
  sequence(s) is coherent.  Also, it is compatible if and only if its
  associated (set of) sequence(s) is compatible and it is consistent
  if and only if its associated (set of) sequence(s) is applicable.
\end{corollary}


\begin{corollary}
  \label{cor:postcondConsCohComp}
  A postcondition is consistent if and only if it is coherent and
  compatible.
\end{corollary}

\begin{figure}[htbp]
  \centering
  \includegraphics[scale = 0.28]{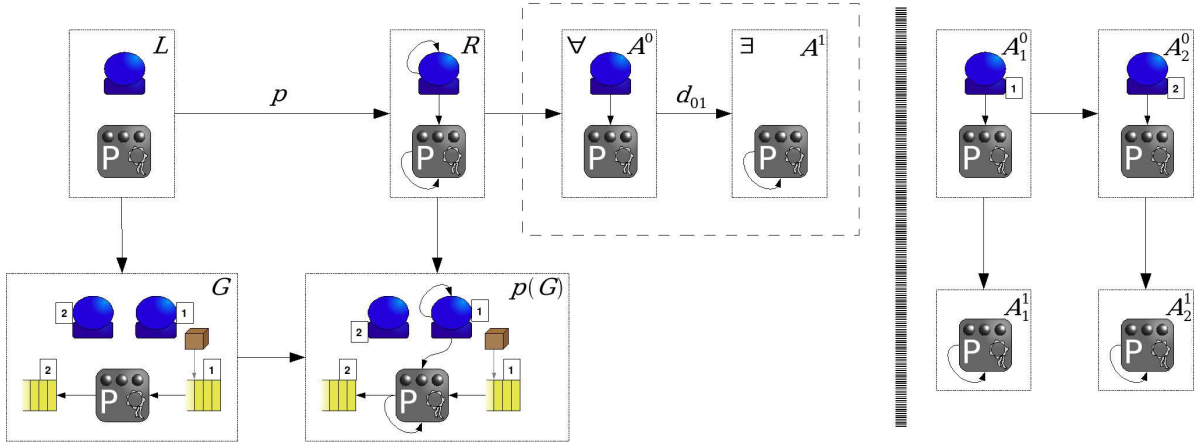}
  \caption{Postcondition Example}
  \label{fig:final_example_GC}
\end{figure}

\noindent \textbf{Example.}$\square$Let's consider the diagram in
Fig.~\ref{fig:final_example_GC} with formula $\exists A^0 \exists A^1
[A^0 \Rightarrow A^1]$. The postcondition states that if an operator
is connected to a machine, such machine is busy. The formula has an
implication so it is not possible to directly generate the set of
sequences because the postcondition also holds when the left of the
implication is false. The closure operator $\widecheck{T}$ reduces the
postcondition to existential quantifiers, which is represented to the
right of the figure. The resulting modified formula would be  $\exists
A^0_1 A^0_2 A^1_1 A^1_2 [ (A^0_1 \Rightarrow A^1_1) \wedge (A^0_2
\Rightarrow A^1_2)]$.

Once the formula has existentials only, we manipulate it to get rid of
implications. Thus, we have $\exists A^0_1 A^0_2 A^1_1 A^1_2
[(\overline{A^0_1} \vee A^1_1) \wedge (\overline{A^0_2} \vee A^1_2)] =
\exists A^0_1 A^0_2 A^1_1 A^1_2 [(\overline{A^0_1} \wedge
\overline{A^1_2}) \vee (\overline{A^0_1} \wedge A^1_2) \vee (A^1_1
\wedge \overline{A^0_2}) \vee (A^1_1 \wedge A^1_2)]$. This leads to a
set of four sequences:
$\{(\overline{id}_{A^0_1}; \overline{id}_{A^0_2}),
(\overline{id}_{A^0_1}; id_{A^1_2}), (id_{A^1_1};
\overline{id}_{A^0_2}), (id_{A^1_1}; id_{A^1_2})\}$. Thus, the graph
$p(G)$ and the production satisfy the postcondition if and only if some
sequence in the set is applicable to $p(G)$. \proofend

Something left undefined is the order of productions $id_{A^i}$ and
$\overline{id}_{A^i}$ in the sequences. Consistency does not depend on
the ordering of productions -- as long as the first to be applied is
production $p$ -- because productions $id$ (and their negation) are
sequentially independent (they do not add nor delete any edge or
node). If they are not sequentially independent then there exists at
least one inconsistency. This inconsistency can be detected using
previous corollaries independently of the order of the productions.

\section{Moving Conditions}
\label{sec:movingConditions}

In this section we give two different proofs that it is possible to
transform preconditions into equivalent postconditions and back
again. The first proof (sketched) makes use of category theory while
the second relies on the characterizations of coherence, G-congruence
and compatibility. To ease exposition we shall focus on the certainty
part only as the nihilation part would follow using the inverse of the
production.


We shall start with a case that can be addressed using
equations~\eqref{eq:postMatch}~--~\eqref{eq:postNAC},
Th.~\ref{th:reductionPost} and Cor.~\ref{cor:equivPostAC_seqs}: When
the transformed postcondition for a given precondition does not
change.\footnote{This is not so unrealistic. For example, if the
  production preserves all elements appearing in the precondition.}
The question of whether it is always possible to transform a
precondition into a postcondition -- and back again -- in this
restricted case would be equivalent to asking for sequential
independence of the production $p$ and the identities $id$ or
$\overline{id}$:
\begin{equation}
  \label{eq:4}
  p;id_{A^n}; \ldots; id_{A^1} = id_{A^n}; \ldots; id_{A^1};p,
\end{equation}
where the sequence to the left of the equality corresponds to a
precondition and the sequence to the right corresponds to its
equivalent postcondition.

\begin{figure}[htb]
  \centering
  \makebox{
    \xymatrix{
      \stackrel{\leftarrow}{A}
      \ar@[blue][ddr]_{m_{\stackrel{\leftarrow}{A}}} 
      \ar@{.>}[rrrr]_{p_A} &&&& \stackrel{\rightarrow}{A}
      \ar@{.>}@[red][ddl]^{m_{\stackrel{\rightarrow}{A}} } & 
      L \ar@[blue][rrr]_{p} \ar@[blue][dd]^{d_L} &&& R
      \ar@{.>}@[red][dd]_{d^{*}_L} &
      \stackrel{\leftarrow}{A} \ar@[blue][rrr]_{p_A}
      \ar@[blue][dd]^{m_{\stackrel{\leftarrow}{A}}} &&&
        \stackrel{\rightarrow}{A}
        \ar@{.>}@[red][dd]_{m_{\stackrel{\rightarrow}{A}}} \\
      & L \ar@[blue][rr]^p \ar@[blue][d]^{m_L} \ar@[blue][ul]_{d_L} &&
      R \ar@{.>}@[red][d]_{m^*_L} \ar@{.>}[ur]^{d^*_L} \\
      & G \ar@{.>}@[red][rr]^{p^*} && H && 
      \stackrel{\leftarrow}{A}
      \ar@{.>}@[red][rrr]^{p_A} &&& \stackrel{\rightarrow}{A} &
      G \ar@{.>}@[red][rrr]^{p_A^*} &&& H
    }
  }
  \caption{Precondition to Postcondition Transformation}
  \label{fig:prePostTrans}
\end{figure}
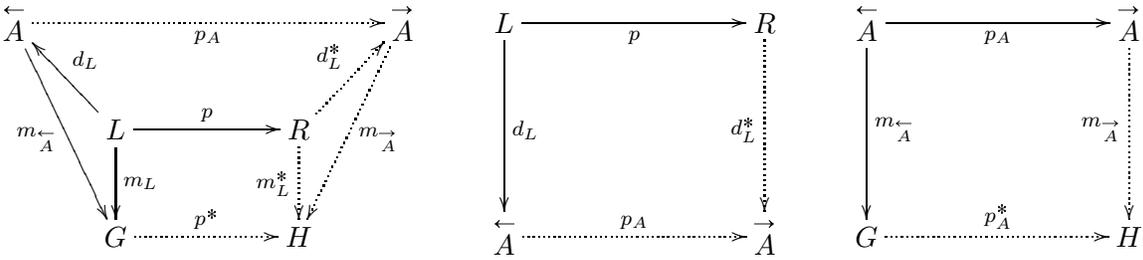

In general the production may act on elements that appear in the
diagram of the precondition, spoiling sequential independence. Left
and center of Fig.~\ref{fig:prePostTrans} -- in which the first basic
AC (match) is considered -- suggest that the pre-to-post
transformation is a categorical pushout\footnote{The square $\left\{L,
    R, \stackrel{\leftarrow}{A}, \stackrel{\rightarrow}{A} \right\}$
  is a pushout where $p$, $L$, $d_L$, $R$ and
  $\stackrel{\leftarrow}{A}$ are known and
  $\stackrel{\rightarrow}{A}$, $p_A$ and $d_L$ need to be calculated.}
in the category of simple digraphs and partial morphisms.

Theorem~\ref{th:reductionPost} proves that any postcondition can be
reduced to the match case. Besides, we can trivially consider total
morphisms (instead of partial ones) by restricting the domain and the
codomain of $p$ to the nodes in $\stackrel{\leftarrow}{A}$. For the
post-to-pre transformation we can either use pullbacks or pushouts
plus the inverse of the production involved.

To see that precondition satisfaction is equivalent to postcondition
satisfaction using category theory, we should check that the different
pushouts can be constructed ($p^*, p_A, p^*_A$, etcetera) and that
$d_L = m_{\stackrel{\leftarrow}{A}} \circ m_L$ and $d^*_L =
m_{\stackrel{\rightarrow}{A}} \circ m^*_L$ (refer to
Fig.~\ref{fig:prePostTrans}). Although some topics remain untouched
such as dangling edges, we shall not carry on with category theory.

\begin{figure}[htbp]
  \centering
  \includegraphics[scale = 0.75]{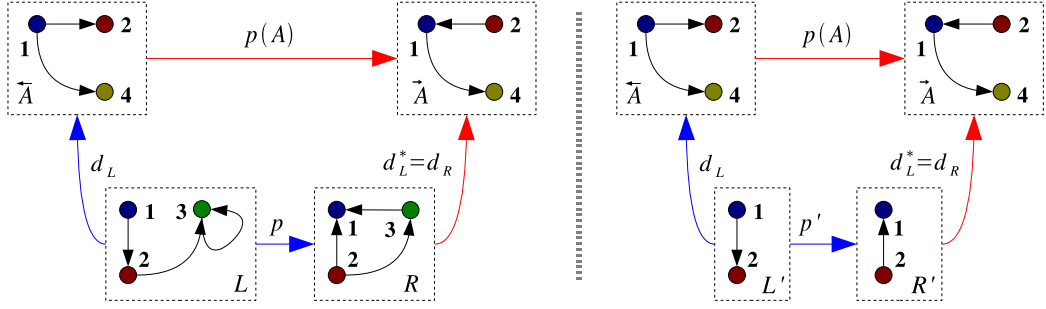}
  \caption{Restriction to Common Parts: Total Morphism}
  \label{fig:pre2PostSimpleEx}
\end{figure}

\noindent \textbf{Example}.$\square$Let be given the precondition
$\stackrel{\leftarrow}{A}$ to the left of
Fig.~\ref{fig:pre2PostSimpleEx} with formula
$\stackrel{\leftarrow}{\mathfrak{f}} = \exists A [A]$. To calculate its
associated postcondition we can apply the production to
$\stackrel{\leftarrow}{A}$ and obtain $\stackrel{\rightarrow}{A}$,
represented also to the left of the same figure. Notice however that
it is not possible to fing a match of $L$ in
$\stackrel{\leftarrow}{A}$ because of node $3$. One possible solution
is to consider $L' = L \, \cap \stackrel{\leftarrow}{A}$ and restrict
the production to those common elements. This is done to the right of
Fig.~\ref{fig:pre2PostSimpleEx} \proofend

\begin{theorem}
  \label{th:prePostPre}
  Any consistent precondition is equivalent to some consistent
  postcondition and vice versa.
\end{theorem}

\noindent \emph{Proof} \\*
$\square$For the post-to-pre transformation roles of $p$ and $p^{-1}$
are interchanged so we shall address only the pre-to-post case. It is
enough to study a single $A$ in the diagram as the same procedure
applies mechanically (Th.~4.1 transforms any precondition into a
sequence of productions). Also, it suffices to state the result for
$id_A$ because $\overline{id}_A$ is similar but the evolution depends
on $p^{-1}$. Finally, we shall assume that $p$ and $id_A$ are not
sequentially independent.

Recall that $G$-congruence guarantees sameness of the initial digraph,
which is what the sequence demands on the host graph. Therefore, all
we have to do is to use $G$-congruence to check the differences in the
two sequences:
\begin{equation}
  \label{eq:1}
  p ; id_{\stackrel{\leftarrow}{A}} \; \longmapsto \;
  id_{\stackrel{\rightarrow}{A}} ; p.
\end{equation}
However, before that we need to guarantee coherence and compatibility
of both sequences (see the hypothesis of Th.~4
in~\cite{MGGCombinatorics}). Coherence gives rise to the following
equation:
\begin{equation}
  \label{eq:3}
  r_{\stackrel{\rightarrow}{A}} R \vee e {\stackrel{\rightarrow}{A}} =
  \mathbf{0} = r\!\!\stackrel{\leftarrow}{A} \vee \,
  e_{\stackrel{\leftarrow}{A}} L \implies
  e\!\!\stackrel{\rightarrow}{A} \vee \, r\!\!\stackrel{\leftarrow}{A}
  \, = \mathbf{0},
\end{equation}
where $e$ and $r$ correspond to $p$, $e_{\stackrel{\leftarrow}{A}}$ to
$id_{\stackrel{\leftarrow}{A}}$ and $r_{\stackrel{\rightarrow}{A}}$ to
$id_{\stackrel{\rightarrow}{A}}$. Fortunately, $id$ is a production
that does nothing, so from the dynamical point of view any conflict
should come from $p$, i.e. $e_{\stackrel{\leftarrow}{A}} =
r_{\stackrel{\rightarrow}{A}} = \mathbf{0}$ which has been used in the
implication of eq.~(\ref{eq:3}).

By consistency we have that $r\!\!\stackrel{\leftarrow}{A} =
\mathbf{0}$ so eq.~(\ref{eq:3}) will be fulfilled if the
postcondition is the precondition but erasing the elements that the
production deletes. A similar reasoning for the nihil part tells us
that we should add to the postcondition all those elements added by
the production.

Compatibility can only be ruined by dangling edges. In~Sec.~6.1
in~\cite{MGGBook} dangling edges are deleted transforming the
production via the opeartor $T_\varepsilon$. This is proved to be
equivalent to defining a sequence by appending a so-called
$\varepsilon$-production. In essence the $\varepsilon$-production just
deletes any dangling edge, thus keeping compatibility. This very same
procedure can be applied now:
\begin{equation}
  \label{eq:7}
  p; id_{\stackrel{\leftarrow}{A}}
  \stackrel{T_\varepsilon}{\longmapsto} p; p_\varepsilon;
  id_{\stackrel{\leftarrow}{A}} \longmapsto p;
  id^{\phantom{i}\varepsilon}_{\stackrel{\leftarrow}{A}}; p_\varepsilon \longmapsto
  id_{\stackrel{\rightarrow}{A}}; p; p_\varepsilon.
\end{equation}

According to Prop.~5 and Th.~4 in~\cite{MGGCombinatorics}, two
compatible and coherent sequences are $G$-congruent if the following
equation (adapted to our case) is fulfilled:
\begin{equation}
  \label{eq:5}
  L_{\stackrel{\leftarrow}{A}} \overline{e} K \left( r \vee
    e_{\stackrel{\leftarrow}{A}} \right) \vee
  K_{\stackrel{\leftarrow}{A}} \overline{r} L (e \vee
  r_{\stackrel{\leftarrow}{A}}) = \mathbf{0}.
\end{equation}
We have that $e_{\stackrel{\leftarrow}{A}} =
r_{\stackrel{\leftarrow}{A}} = \mathbf{0}$. Also,
$K_{\stackrel{\leftarrow}{A}} = \mathbf{0}$ because
$id_{\stackrel{\leftarrow}{A}}$ acts on the certainty part and
$\overline{e}r = r$ (see e.g. Prop.~4.1.4 in~\cite{handbook}). We are
left with
\begin{equation}
  \label{eq:6}
  L_{\stackrel{\leftarrow}{A}} K = K\!\!\stackrel{\leftarrow}{A} =
  \mathbf{0},
\end{equation}
which is guaranteed by compatibility: once
$id_{\stackrel{\leftarrow}{A}}$ is transformed into
$id^{\phantom{i}\varepsilon}_{\stackrel{\leftarrow}{A}}$ in
eq.~(\ref{eq:7}) there can not be any potential dangling edge, except
those to be deleted by $p$ in the last step. \proofend 

It is worth stressing the fact that the transformation between pre and
postconditions preserve consistency of the application condition. We
have seen in this section that $p$ not only acts on $L$ but on the
whole precondition. We can therefore extend the notation:
\begin{equation}\label{eq:prePostTransFuncNotation}
  \stackrel{\rightarrow}{A} \, = p (\stackrel{\leftarrow}{A}), \qquad
  \stackrel{\rightarrow}{A} \, = \langle\ 
    \!\!\stackrel{\leftarrow}{A}, p \rangle.
\end{equation}

Pre-to-post and post-to-pre transformations can affect the diagram and
the formula. See the example below. There are two clear cases:
\begin{itemize}
\item The application condition requires the graph to appear and the
  production deletes all its elements.
\item The application condition requires the graph not to appear and
  the production adds all its elements.
\end{itemize}

For a given application condition AC it is not necessarily true that
$A = p^{-1};p(A)$ because some new elements may be added and some
obsolete elements discarded.  What we will get is an equivalent
condition adapted to $p$ that holds whenever $A$ holds and fails to be
true whenever $A$ is false.

\begin{figure}[htbp]
  \centering
  \includegraphics[scale = 0.75]{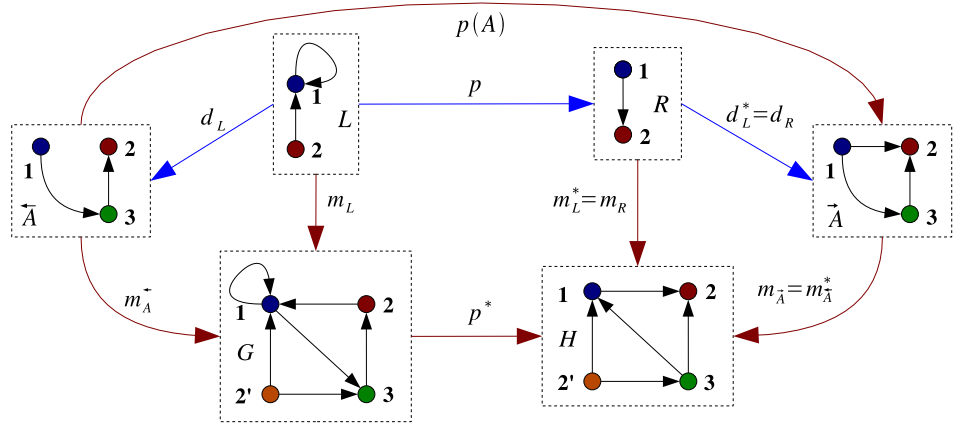}
  \caption{Precondition to Postcondition Example}
  \label{fig:pre2PostEx}
\end{figure}

\noindent\textbf{Example}.$\square$In Fig.~\ref{fig:pre2PostEx} there
is a very simple transformation of a precondition into a postcondition
through morphism $p(A)$. The associated formula to the precondition
$\stackrel{\leftarrow}{A}$ that we shall consider is
$\stackrel{\leftarrow}{\mathfrak{f}} = \exists A [A]$. The production
deletes two arrows and adds a new one. The overall effect is reverting
the direction of the edge between nodes $1$ and $2$ and deleting the
self-loop in node $1$. Notice that $m_L$ can not match node $2$ to
$2'$ in $G$ because of the edge $(3,2)$ in the application condition.

Suppose we had a (redundant) graph $B$ made up of a single node $1$
with a self loop in the precondition and with formula
$\stackrel{\leftarrow}{\mathfrak{f}} = \exists AB [AB]$. The formula
in the postcondition would still be
$\stackrel{\rightarrow}{\mathfrak{f}} = \exists A [A]$.

The opposite transformation, from postcondition into precondition, can
be obtained by reverting the arrow, i.e. through $p^{-1}(A)$.  More
general schemes can be studied applying the same principles.

Let $\mathcal{A} = p^{-1} \circ p \left( \stackrel{\leftarrow}{A}
\right)$. If a pre-post-pre transformation is carried out, we will
have $\stackrel{\leftarrow}{A} \,\neq \mathcal{A}$ because edge (2,1)
would be added to $\stackrel{\leftarrow}{A}$. However, it is true that
$\mathcal{A} = p^{-1}\circ p \left( \mathcal{A} \right)$.

Note that in fact $id_{\stackrel{\leftarrow}{A}}$ and $p$ are
sequentially independent if we limit ourselves to edges, so it would
be possible to simply move the precondition to a postcondition as it
is. Nonetheless, we have to consider nodes 1 and 2 as the common parts
between $L$ and $\stackrel{\leftarrow}{A}$. This is the same kind of
restriction as the one illustrated in
Fig.~\ref{fig:pre2PostSimpleEx}. \proofend

If the pre-post-pre transformation is thought of as an operator $T_p$
acting on application conditions, then it fulfills
\begin{equation}
  \label{eq:2}
  T_p^2 = id,
\end{equation}
where $id$ is the identity. The same would also be true for a
post-pre-post transformation.

A possible interpretation of eq.~\eqref{eq:2} is that the definition
of the application condition can vary from the \emph{natural} one,
according to the production under consideration. Pre-post-pre or
post-pre-post transformations adjust application conditions to the
corresponding production.

When defining diagrams some ``practical problems'' may turn up.  For
example, if the diagram $\mathfrak{d} = \left( L
  \stackrel{d_{L0}}{\rightarrow} A^0 \stackrel{d_{10}}{\leftarrow} A^1
\right)$ is considered then there are two potential problems:
\begin{enumerate}
\item The direction in the arrow $A^0 \leftarrow A^1$ is not the
  natural one.  Nevertheless, injectiveness allows us to safely revert
  the arrow, $d_{01} = d^{-1}_{10}$.
\item Even though we only formally state $d_{L0}$ and $d_{10}$, other
  morphisms naturally appear and need to be checked out, e.g. $d_{L1}:
  R \rightarrow A^1$.  New morphisms should be considered if they
  relate at least one element.\footnote{Otherwise stated: Any
    condition made up of $n$ graphs $A^i$ can be identified as the
    complete graph $K_n$, in which nodes are graphs $A^i$ and
    morphisms are $d_{ij}$.  Whether this is a directed graph or not
    is a matter of taste (morphisms are injective).}
\end{enumerate}

\section{Delocalization and Variable Nodes}
\label{sec:delocalization}

In this section we touch on \emph{delocalization} of graph constraints
and application conditions as well as their equivalence. Also, we
shall pave the way to multidigraph rewriting to be studied in detail
in Sec.~\ref{sec:fromSimpleDigraphsToMultidigraphs}.

Let $s = p_n; \ldots; p_1$ be a sequence of productions with their
corresponding ACs. We have seen in Th.~\ref{th:reductionPost} that
preconditions and postconditions are equivalent and in
Th.~\ref{th:prePostPre} that they can be transformed into sequences of
productions. As a precondition in $p_{i+1}$ is the same as a
postcondition in $p_i$, we see that ACs can be moved arbitrarily
inside a sequence.

Similarly, constraints set on the intermediate states of a derivation
can be moved among them. A graph constraint GC set in the initial
state $G$ to which a production $p$ is going to be applied is
equivalent to the precondition
\begin{equation}
  \label{eq:8}
  \mathfrak{f}_{pre} = \exists L \exists K \left[ L \wedge P\left( K,
      \overline{G} \right) \wedge \mathfrak{f}_{GC}\right].
\end{equation}
If the GC is set on the final state $ H = p(G)$ to which the
production $p$ has been applied, there is an equivalent postcondition:
\begin{equation}
  \label{eq:9}
  \mathfrak{f}_{post} = \exists R \exists Q \left[ P(R, H) \wedge P
    \left( Q, \overline{H} \right) \wedge \mathfrak{f}_{GC}\right].
\end{equation}
In both cases the diagrams are given by the LHS or the RHS plus the
diagram of the graph constraint. We call this property of application
conditions and graph constraints \emph{delocalization}.

We shall now address variable nodes which will be used to enhance MGG
functionality to deal with multidigraphs. Graph transformation with
variables is studied in~\cite{Hof05}. We shall summarize the proposal
in~\cite{Hof05} and propound an alternative way to close the section.

If instead of nodes of fixed type variable, types are allowed we get a
so called \emph{graph pattern}. A \emph{rule scheme} is just a
production in which graphs are graph patterns. A \emph{substitution
  function} $\iota$ specifies how variable names taking place in a
production are substituted.  A rule scheme $p$ is instantiated via
substitution functions producing a particular production. For example,
for substitution function $\iota$ we get $p^{\,\iota}$. The set of
production instances for $p$ is defined as the set $\mathcal{I}(p) =
\left\{ p^{\,\iota} \; \vert \right.$ $\iota$ is a
substitution$\left.\right\}$. The \emph{kernel} of a graph $G$,
$ker(G)$, is defined as the graph resulting when all variable nodes
are removed.  It might be the case that $ker(G) = \emptyset$.

The basic idea is to reduce any rule scheme to a set of rule
instances.  Note that it is not possible in general to generate
$\mathcal{I}(p)$ because this set can be infinite.  The way to proceed
is not difficult:
\begin{enumerate}
\item Find a match for the kernel of $L$.
\item Induce a substitution $\iota$ such that the match for the kernel
  becomes a full match $m:L^{\iota} \rightarrow G$.
\item Construct the instance $R^{\,\iota}$ and apply $p^{\,\iota}$ to
  get the direct derivation $G \stackrel{p^{\,\iota}}{\Longrightarrow}
  H$.
\end{enumerate}

As an alternative, we may extend the concept of type
assignment. Recall from Sec.~\ref{sec:MGGs} that types are assigned by
a function from the set of nodes $|V|$ of a simple digraph $G$ to some
fixed set $T$ of types, $\lambda:|V| \to T$. Instead, we shall define
\begin{equation}
  \label{eq:10}
  \lambda:|V| \longrightarrow \mathcal{P}(T)\backslash\emptyset,
\end{equation}
where $\mathcal{P}(T)$ is the power set\footnote{The set of all
  subsets.} of $T$ except for the empty set because we do not permit
nodes without types.

When two matrices are operated, the types of a fixed node will be the
intersection of the nodes operated. For example, suppose that we
\textbf{and} two matrices $C = AB$ and that the (set of) nodes
associated to the elements $a$ and $b$ are $\lambda(a)$ and
$\lambda(b)$, respectively. Then, $\lambda(c) = \lambda(a) \cap
\lambda(b)$. The operation would not be allowed in case $\lambda(c) =
\lambda(a) \cap \lambda(b) = \emptyset$. 

\begin{figure}[htbp]
  \centering
  \includegraphics[scale = 0.8]{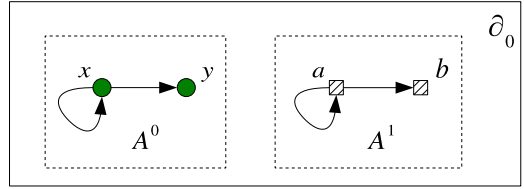}
  \caption{Example of Graph Constraint}
  \label{fig:simplifiedDiagram2}
\end{figure}

\noindent \textbf{Example}.$\square$ Let a type of nodes be
represented by squares (call them \emph{multinodes}) and the rest
(call them \emph{simple nodes}) by colored circles. The set of types
$T$ is split into two: multinodes and simple nodes.

Let's consider the graph constraint $GC_0 = \left( \mathfrak{d}_0,
  \mathfrak{f}_0 \right)$, with $\mathfrak{d}_0$ the diagram depicted
in Fig.~\ref{fig:simplifiedDiagram2} made up of the graphs $A^0$ and
$A^1$, along with the formula $\mathfrak{f}_0 = \forall A^0 A^1[
\overline{Q}(A^0) \overline{Q}(A^1)]$. This graph constraint is
``edges must connect nodes and multinodes alternatively but no edge
is allowed to be incident to two multinodes or to two simple nodes,
including self-loops''.

In the graph $A^0$ of Fig.~\ref{fig:simplifiedDiagram2}, $x$ and $y$
represent variable nodes while $a$ and $b$ in $A^1$ have a fixed
type. We may think of $a$ and $b$ as variable nodes whose set of types
has a single element. \proofend


\section{From Simple Digraphs to Multidigraphs}
\label{sec:fromSimpleDigraphsToMultidigraphs}

In this section we show how MGG can deal with multidigraphs (directed
graphs allowing multiple parallel edges) just by considering variable
nodes. At first sight this might seem a hard task as MGG heavily
depends on adjacency matrices.  Adjacency matrices are well suited for
simple digraphs but can not cope with parallel edges. This section can
be thought of as a \emph{theoretical application} of graph constraints
and application conditions to Matrix Graph Grammars.

The idea is not difficult: A special kind of node (call it
\emph{multinode} in contrast to \emph{simple node}) associated to
every edge in the graph is introduced, i.e. edges in the multidigraph
are substituted by multinodes in a simple digraph representation of
the multidigraph.  Graphically, multinodes will be represented by a
filled square while normal nodes will appear as colored circles. See
the example by the end of Sec.~\ref{sec:delocalization}


Operations previously specified on edges now act on multinodes: Adding
an edge is transformed into a multinode addition and edge deletion
becomes multinode deletion.  There are edges that link multinodes to
their source and target simple nodes.

Some restrictions (application conditions) to be imposed on the
actions that can be performed on multinodes exist, as well as on the
shape or topology of permitted graphs (graph constraints). Not every
possible graph with multinodes represents a multidigraph.

\begin{figure}[htbp]
  \centering
  \includegraphics[scale = 0.9]{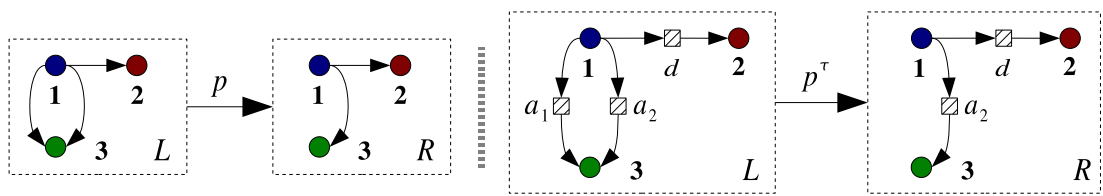}
  \caption{Multidigraph with Two Outgoing Edges}
  \label{fig:simpleMultiGraph}
\end{figure}

\noindent\textbf{Example}.$\square$Consider the simple production in
Fig.~\ref{fig:simpleMultiGraph} with two parallel edges between nodes
$1$ and $3$.  As commented above, multinodes are represented by square
nodes while normal nodes are left unchanged.  When $p$ deletes an
edge, $p^\tau$ deletes a multinode.  Adjacency matrices for $p^\tau$
are:

\begin{align*}
  L = \left[
    \begin{array}{ccccccc}
      \vspace{-10pt}
      0 & 0 & 0 & 1 & 1 & 1 & \vert\;\; 1_{\phantom{2}} \\
      \vspace{-10pt}
      0 & 0 & 0 & 0 & 0 & 0 & \vert\;\; 2_{\phantom{2}} \\
      \vspace{-10pt}
      0 & 0 & 0 & 0 & 0 & 0 & \vert\;\; 3_{\phantom{2}} \\
      \vspace{-10pt}
      0 & 0 & 1 & 0 & 0 & 0 & \vert\;\; a_1 \\
      \vspace{-10pt}
      0 & 0 & 1 & 0 & 0 & 0 & \vert\;\; a_2 \\
      \vspace{-10pt}
      0 & 1 & 0 & 0 & 0 & 0 & \vert\;\; d_{\phantom{2}} \\
      \vspace{-10pt}
    \end{array} \right] \qquad
  R &= \left[
    \begin{array}{cccccc}
      \vspace{-10pt}
      0 & 0 & 0 & 1 & 1 & \vert\;\; 1_{\phantom{2}} \\
      \vspace{-10pt}
      0 & 0 & 0 & 0 & 0 & \vert\;\; 2_{\phantom{2}} \\
      \vspace{-10pt}
      0 & 0 & 0 & 0 & 0 & \vert\;\; 3_{\phantom{2}} \\
      \vspace{-10pt}
      0 & 0 & 1 & 0 & 0 & \vert\;\; a_2 \\
      \vspace{-10pt}
      0 & 1 & 0 & 0 & 0 & \vert\;\; d_{\phantom{2}} \\
      \vspace{-10pt}
    \end{array} \right] \\
  K = \left[
    \begin{array}{ccccccc}
      \vspace{-10pt}
      0 & 0 & 0 & 0 & 0 & 0 & \vert\;\; 1_{\phantom{2}} \\
      \vspace{-10pt}
      0 & 0 & 0 & 1 & 0 & 0 & \vert\;\; 2_{\phantom{2}} \\
      \vspace{-10pt}
      0 & 0 & 0 & 1 & 0 & 0 & \vert\;\; 3_{\phantom{2}} \\
      \vspace{-10pt}
      1 & 1 & 0 & 1 & 1 & 1 & \vert\;\; a_1 \\
      \vspace{-10pt}
      0 & 0 & 0 & 1 & 0 & 0 & \vert\;\; a_2 \\
      \vspace{-10pt}
      0 & 0 & 0 & 1 & 0 & 0 & \vert\;\; d_{\phantom{2}} \\
      \vspace{-10pt}
    \end{array} \right] \qquad
  \;\;e &= \left[
    \begin{array}{ccccccc}
      \vspace{-10pt}
      0 & 0 & 0 & 1 & 0 & 0 & \vert\;\; 1_{\phantom{2}} \\
      \vspace{-10pt}
      0 & 0 & 0 & 0 & 0 & 0 & \vert\;\; 2_{\phantom{2}} \\
      \vspace{-10pt}
      0 & 0 & 0 & 0 & 0 & 0 & \vert\;\; 3_{\phantom{2}} \\
      \vspace{-10pt}
      0 & 0 & 1 & 0 & 0 & 0 & \vert\;\; a_1 \\
      \vspace{-10pt}
      0 & 0 & 0 & 0 & 0 & 0 & \vert\;\; a_2 \\
      \vspace{-10pt}
      0 & 0 & 0 & 0 & 0 & 0 & \vert\;\; d_{\phantom{2}} \\
      \vspace{-10pt}
    \end{array} \right]  \nonumber
\end{align*}


In a real situation, a development tool such as AToM$^3$ or
AGG\footnote{\url{http://moncs.cs.mcgill.ca/MSDL/research/projects/AToM3/}
  for AToM$^3$ and \url{http://www.gratra.org/} for AGG and some other
  tools.} should take care of all these representation issues.  A user
would see what appears to the left of Fig.~\ref{fig:simpleMultiGraph}
and not what is depicted to the right of the same figure. \proofend

Some restrictions on what a production can do to a multidigraph are
necessary in order to obtain a multidigraph again. Think for example
the case in which after applying some production we get a graph in
which there is an isolated multinode (which would stand for an edge
with no source nor target nodes). All we have to do is to find the
properties that define one edge and impose them on multinodes as graph
constraints:
\begin{enumerate}
\item A simple node (resp., multinode) can not be directly connected
  to another simple node (resp., multinode).
\item Edges (encoded as multinodes) always have a simple node as
  source and a simple node as target.
\end{enumerate}

\begin{figure}[htbp]
  \centering
  \includegraphics[scale = 0.9]{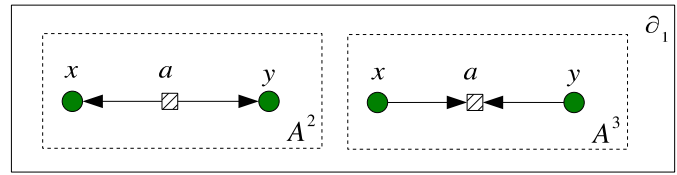}
  \caption{Multidigraph Constraints}
  \label{fig:multidigraphGC2}
\end{figure}

First condition above is addressed in the example of
Sec.~\ref{sec:delocalization} with graph constraint $GC_0$. See
Fig.~\ref{fig:simplifiedDiagram2}. The second condition can be encoded
as another graph constraint $GC_1 = \left( \mathfrak{d}_1,
  \mathfrak{f}_1 \right)$. The diagram can be found in
Fig.~\ref{fig:multidigraphGC2} and the formula is $\mathfrak{f}_1 =
\forall A^2 A^3 [ \overline{A^2} \; \overline{A^3} ]$.

\begin{theorem}
  \label{th:multidigraph}
  Any multidigraph is isomorphic to some simple digraph $G$ together
  with the graph constraint $MC = \left( \mathfrak{d}_0 \cup
    \mathfrak{d}_1, \mathfrak{f}_0 \wedge \mathfrak{f}_1 \right)$.
\end{theorem}

\noindent \emph{Proof (sketch)}\\*
$\square$A graph with multiple edges $M = \left( V,E,s,t \right)$
consists of disjoint finite sets $V$ of nodes and $E$ of edges and
source and target functions $s:E \rightarrow V$ and $t:E \rightarrow
V$, respectively.  Function $v = s(e)$, $v \in V$, $e \in E$ returns
the node source $v$ for edge $e$.  We are considering multidigraphs
because the pair function $(s,t):E\rightarrow V\times V$ need not be
injective, i.e. several different edges may have the same source and
target nodes.  We have digraphs because there is a distinction between
source and target nodes.  This is the standard definition found in any
textbook.

It is clear that any $M$ can be represented as a multidigraph $G$
satisfying $MC$.  The converse also holds.  To see it, just consider
all possible combinations of two nodes and two multinodes and check
that any problematic situation is ruled out by $MC$. Induction
finishes the proof.\proofend

The multidigraph constraint $MC$ must be fulfilled by any host graph.
If there is a production $p:L \rightarrow R$ involved, $MC$ has to be
transformed into an application condition over $p$.  In fact, the
multidigraph constraint should be demanded both as precondition and
postcondition. This is easily achieved by means of eqs.~(\ref{eq:8})
and~(\ref{eq:9}).

This section is closed analyzing what behavior we have for
multidigraphs with respect to dangling edges.  With the theory as
developed so far, if a production specifies the deletion of a simple
node then an $\varepsilon$-production would delete any edge incident
to this simple node, connecting it to any surrounding multinode.  But
restrictions imposed by MC do not allow this so any production with
potential dangling edges can not be applied.

In order to automatically delete any potential multiple dangling edge,
$\varepsilon$-productions need to be restated by defining them at a
multidigraph level, i.e.  $\varepsilon$-productions have to delete any
potential ``dangling multinode''.  \index{x@$\Xi$-production}A new
type of productions ($\Xi$-productions) are introduced to get rid of
annoying edges\footnote{Edges connect simple nodes and multinodes.}
that would dangle when multinodes are also deleted by
$\varepsilon$-productions. We will not develop the idea in detail and
will limit to describe the concepts.  The way to proceed is to define
the appropriate operator $T_\Xi$ and redefine the operator
$T_\varepsilon$.

\begin{figure}[htbp]
  \centering
  \includegraphics[scale = 0.73]{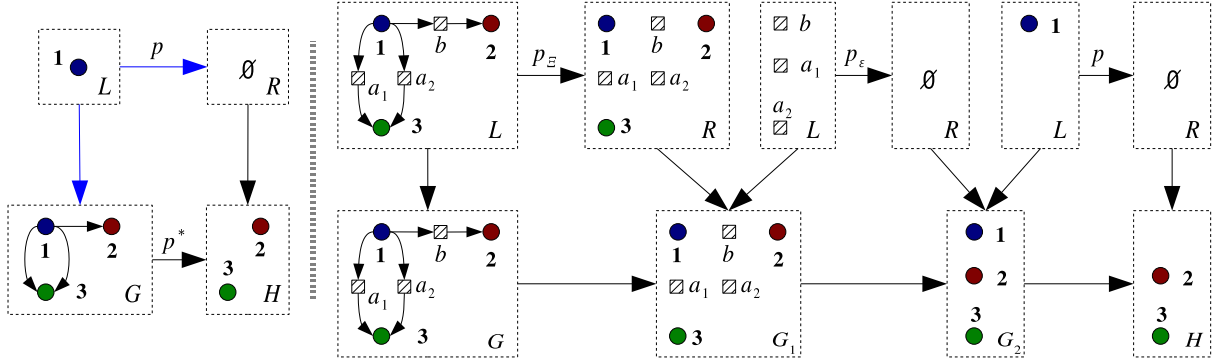}
  \caption{$\varepsilon$-production and $\Xi$-production}
  \label{fig:spoMultidigraph}
\end{figure}

A production $p:L \rightarrow R$ between multidigraphs that deletes
one simple node $n_1$ may give rise to one $\varepsilon$-production
that deletes one or more multinodes $m_i$ (those ``incident'' to $n_1$
not deleted by the grammar rule).  This $\varepsilon$-production can
in turn be applied only if any edge incident to the $m_i$'s has
already been erased, hence possibly provoking the appearance of one
$\Xi$-production.

This process is depicted in Fig.~\ref{fig:spoMultidigraph} where, in
order to apply production $p$, productions $p_\varepsilon$ and $p_\Xi$
need to be applied in advance
\begin{equation}
  p \longmapsto p; p_\varepsilon; p_\Xi.
\end{equation}

Eventually, one could simply compose the $\Xi$-production with its
$\varepsilon$-production, renaming it to $\varepsilon$-production and
defining it as the way to deal with dangling edges in case of multiple
edges, fully recovering the standard behavior in MGG. As commented
above, a potential user of a development tool such as AToM$^3$ would
still see things as in the simple digraph case, with no need to worry
about $\Xi$-productions.

Another theoretical use of application conditions and graph
constraints is the encoding of Turing Machines and Boolean
Circuits using Matrix Graph Grammars (see~\cite{MGGmodel}). However,
they are not necessary for Petri nets (see Chap.~10
in~\cite{MGGBook}).

\section{Conclusions and Future Work}
\label{sec:conclusions}

In the present contribution we have introduced preconditions and
postconditions for MGGs, proving that there is an equivalent set of
sequences of plain rules to any given postcondition. Besides,
coherence, compatibility and consistency of postconditions have been
characterized in terms of already known concepts for sequences. We
have also proved that it is always possible to transform any
postcondition into an equivalent precondition and vice
versa. Moreover, we have seen that restrictions are delocalized if a
sequence is under consideration. An alternative way to that
in~\cite{Hof05} to tackle variable nodes has also been proposed. This
allows us to extend MGG to cope with multidigraphs without major
modifications to the theory.

In~\cite{MGGfundamenta} there is an exhaustive comparison of the
application conditions in MGG and other proposals. The main papers to
the best of our knowledge that tackle this topic are~\cite{AC:Ehrig}
(with the definition of ACs), \cite{Habel, HP09} where GCs and ACs are
extended with nesting and satisfiability, and also~\cite{Rensink} in
which ACs are generalized to arbitrary levels of nesting (though
restricted to trees).

For future work, we shall generalize already studied concepts in MGG
for multidigraphs such as coherence, compatibility, initial digraphs,
graph congruence, reachability, etcetera. Our main interest, however,
will be focused on complexity theory and the application of MGG to the
study of complexity classes, \textbf{P} and \textbf{NP} in
particular. \cite{MGGmodel} follows this line of research.



\begin{thebibliography}{10}\label{bibliography}

\bibitem{AGG} AGG, The Attributed Graph Grammar system. {\texttt
    {http://tfs.cs.tu-berlin.de/agg/}}.

\bibitem{braket} Bra-ket notation intro:
  {\texttt{http://en.wikipedia.org/wiki/Bra-ket\_notation}}

\bibitem{DPO:handbook} Corradini, A., Montanari, U., Rossi, F., Ehrig,
  H., Heckel, R., L\"owe, M. 1999.  {\em Algebraic Approaches to Graph
    Transformation - Part I: Basic Concepts and Double Pushout
    Approach}.  In~\cite{handbook}, pp.: 163-246

\bibitem{Courcelle} Courcelle, B. 1997. \emph{The expression of graph
    properties and graph transformations in monadic second-order
    logic}.  In~\cite{handbook}, pp.: 313-400.


\bibitem{SPO:handbook} Ehrig, H., Heckel, R., Korff, M., L\"owe, M.,
  Ribeiro, L., Wagner, A., Corradini, A. 1999.  {\em Algebraic
    Approaches to Graph Transformation - Part II: Single Pushout
    Approach and Comparison with Double Pushout Approach.}
  In~\cite{handbook}, pp.: 247-312.

\bibitem{AC:Ehrig} Ehrig, H., Ehrig, K., Habel, A., Pennemann, K.-H.
  {\em Constraints and Application Conditions: From Graphs to
    High-Level Structures}.  Proc. ICGT'04. LNCS 3256, pp.: 287-303.
  Springer.

\bibitem{graGraBook} Ehrig, H., Ehrig, K., Prange, U., Taentzer, G.
  2006.  {\em Fundamentals of Algebraic Graph Transformation.}
  Springer.


\bibitem{Habel} Habel, A., Pennemann, K.-H. 2005. {\em Nested
    Constraints and Application Conditions for High-Level Structures}.
  In Formal Methods in Software and Systems Modeling, LNCS 3393, pp.
  293-308. Springer.

\bibitem{HP09} Habel, A., Penneman, K.-H. 2009. {\em Correctness of High-Level
Transformation Systems Relative to Nested
Conditions}. Math. Struct. Comp. Science 19(2), pp.: 245--296.

\bibitem{Heckel} Heckel, R., K\"uster, J.-M-., Taentzer, G. 2002.
  {\em Confluence of typed attributed graph transformation systems}.
  Proc. ICGT'02, LNCS 2505, pp. 161--176. Springer.

\bibitem{HeckelW95} Heckel, R., Wagner, A. 1995.  {\em Ensuring
    consistency of conditional graph rewriting - a constructive
    approach.}, Electr. Notes Theor. Comput. Sci. (2).

\bibitem{Hof05} Hoffman, B. 2005.  \emph{Graph Transformation with
    Variables}.  In Graph Transformation, Vol. 3393/2005 of LNCS, pp.
  101-115. Springer.

\bibitem{Lambers} Lambers, L., Ehrig, H., Orejas, F. 2006.  {\em
    Conflict Detection for Graph Transformation with Negative
    Application Conditions}.  Proc ICGT'06, LNCS 4178, pp.: 61-76.
  Springer.

\bibitem{JVLC} de Lara, J., Vangheluwe, H. 2004.  {\em Defining Visual
    Notations and Their Manipulation Through Meta-Modelling and Graph
    Transformation}.  Journal of Visual Languages and Computing.
  Special section on ``Domain-Specific Modeling with Visual
  Languages'', Vol 15(3-4), pp.: 309-330. Elsevier Science.

\bibitem{JuanPP_1} P\'erez Velasco, P. P., de Lara, J. 2006.
  \emph{Towards a New Algebraic Approach to Graph Transformation: Long
    Version.}  Tech. Rep. of the School of Comp. Sci., Univ.
  Aut\'onoma Madrid.
  {\texttt{http://www.ii.uam.es/$\sim$jlara/investigacion/techrep\_03\_06.pdf}}.

\bibitem{JuanPP_2} P\'erez Velasco, P. P., de Lara, J. 2006.  {\em
    Matrix Approach to Graph Transformation: Matching and Sequences}.
  Proc ICGT'06, LNCS 4178, pp.:122-137. Springer.

\bibitem{JuanPP_4} P\'erez Velasco, P. P., de Lara, J. 2007.  {\em
    Using Matrix Graph Grammars for the Analysis of Behavioural
    Specifications: Sequential and Parallel Independence} Proc.
  PROLE'07, pp.: 11-26. Electr. Notes
  Theor. Comput. Sci. (206). pp.:133--152. Elsevier.

\bibitem{GTVC} P\'erez Velasco, P. P., de Lara, J. 2007.  {\em
    Analysing Rules with Application Conditions using Matrix Graph
    Grammars}.  Graph Transformation for Verification and Concurrency
  (GTVC) workshop.

\bibitem{MGGBook} P\'erez Velasco, P. P. 2009.  \emph{Matrix Graph
    Grammars: An Algebraic Approach to Graph Dynamics}.  ISBN
  978-3639212556. VDM Verlag. Also available as e-book at:
  {\texttt{http://www.mat2gra.info/}} and \texttt{arXiv:0801.1245v1}.

\bibitem{MGGCombinatorics} P\'erez Velasco, P. P., de Lara, J. 2009.
  \emph{A Reformulation of Matrix Graph Grammars with Boolean
    Complexes}. The Electronic Journal of Combinatorics. Vol
  16(1). R73. Available at: {\texttt{http://www.combinatorics.org/}}.

\bibitem{MGGmodel} P\'erez Velasco, P. P. 2009. \emph{Matrix Graph
    Grammars as a Model of Computation}. Preliminary version available
  at {\texttt{arXiv:0905.1202v2}}.

\bibitem{MGGfundamenta} P\'erez Velasco, P. P., de Lara,
  J. 2010. \emph{Matrix Graph Grammars with Application
    Conditions}. To appear in Fundamenta Informaticae. Also available
  at {\texttt{arXiv:0902.1809v2}}.

\bibitem{Rensink} Rensink, A. 2004. {\em Representing First-Order
    Logic Using Graphs.}  Proc. ICGT'04, LNCS 3256, pp.: 319-335.
  Springer.

\bibitem{handbook} {Rozenberg, G.} 1997.  {\em Handbook of Graph
    Grammars and Computing by Graph Transformation. Vol 1. } World
  Scientific.

\end{thebibliography}
\end{document}